\documentclass[preprint,12pt]{elsarticle}

\usepackage{amssymb}
\usepackage{amsmath}

\usepackage{graphicx}
\begin{document}

\begin{frontmatter}



\title{CRepair: CVAE-based Automatic Vulnerability Repair Technology}


\author[label1]{Penghui Liu}

\author[label1]{Yingzhou Bi\corref{cor1}}
\author[label1]{Jiangtao Huang}
\author[label1]{Xinxin Jiang}
\author[label1]{Lianmei Wang}

\cortext[cor1]{Corresponding author. Email: byz@nnnu.edu.cn}

\affiliation[label1]{organization={Nanning Normal University},
	addressline={Address One}, 
	city={Nanning},
	postcode={530110}, 
	state={Guangxi Zhuang Autonomous Region},
	country={China}}

\begin{abstract}
Software vulnerabilities are flaws in computer software systems that pose significant threats to the integrity, security, and reliability of modern software and its application data. These vulnerabilities can lead to substantial economic losses across various industries. Manual vulnerability repair is not only time-consuming but also prone to errors. To address the challenges of vulnerability repair, researchers have proposed various solutions, with learning-based automatic vulnerability repair techniques gaining widespread attention. However, existing methods often focus on learning more vulnerability data to improve repair outcomes, while neglecting the diverse characteristics of vulnerable code, and suffer from imprecise vulnerability localization.To address these shortcomings, this paper proposes CRepair, a CVAE-based automatic vulnerability repair technology aimed at fixing security vulnerabilities in system code. We first preprocess the vulnerability data using a prompt-based method to serve as input to the model. Then, we apply causal inference techniques to map the vulnerability feature data to probability distributions. By employing multi-sample feature fusion, we capture diverse vulnerability feature information. Finally, conditional control is used to guide the model in repairing the vulnerabilities.Experimental results demonstrate that the proposed method significantly outperforms other benchmark models, achieving a perfect repair rate of 52\%. The effectiveness of the approach is validated from multiple perspectives, advancing AI-driven code vulnerability repair and showing promising applications.
\end{abstract}

\begin{graphicalabstract}
\end{graphicalabstract}

\begin{highlights}
\item Research highlight 1
\item Research highlight 2
\end{highlights}


\end{frontmatter}



\section{INTRODUCTIO}
\label{sec1}
Software vulnerabilities refer to flaws in the computational logic or code of a software system, which can be exploited by malicious actors to cause significant harm to the system\cite{1}. With the rapid development of the internet, both the complexity of software and the number of vulnerabilities have surged, becoming a major threat to cybersecurity. Traditional vulnerability repair methods require extensive human intervention, are inefficient, and often lead to new vulnerabilities\cite{2}. Due to the diverse root causes and repair methods for different vulnerabilities, researchers have classified vulnerabilities based on their intrinsic characteristics to facilitate understanding and repair. This has led to the formation of various vulnerability types. Currently, the international Common Weakness Enumeration (CWE) system includes 939 types of vulnerabilities\cite{3}, aiding researchers in vulnerability analysis and remediation. However, the vast number and complexity of vulnerability types make it difficult to analyze repair mechanisms for each one. As a result, general vulnerability repair methods, which can be applied to multiple types of vulnerabilities, have garnered significant attention from researchers.The current mainstream approach for general vulnerability repair is a history-driven deep learning model. This method takes vulnerable code as input and repair patches as output, focusing on learning from historical repair versions to achieve automatic vulnerability repair. However, existing public vulnerability datasets are often small in size and low in quality, and vulnerable code is typically lengthy, making both vulnerability localization and repair challenging. The key challenge for automatic vulnerability repair lies in how to use limited datasets to accurately repair vulnerabilities and quickly locate them.

\par
To address the issue of limited vulnerability datasets, Nong et al.\cite{4} attempted to generate realistic vulnerabilities by injecting them into real code data using neural code techniques. However, there remains a significant difference between synthetic data and real vulnerability data. To focus on the repair of actual vulnerability data, Chen et al.\cite{5} proposed VRepair, an automatic vulnerability repair technique based on transfer learning. VRepair collects vulnerability repair data by analyzing GitHub events, ultimately filtering and constructing a corpus of 650,000 pairs of C language repair examples for pre-training. It then fine-tunes on the CVEFixes and Big-Vul vulnerability datasets, mitigating the issue of limited vulnerability data and validating the effectiveness of transfer learning in addressing data scarcity.However, its pretraining approach is relatively simplistic, failing to capture the diverse intrinsic characteristics of vulnerabilities.To overcome this limitation, Fu et al.\cite{6} introduced VulRepair, an automatic vulnerability repair technique based on the T5 model. VulRepair applies four pretraining strategies across 8.35 million functions, resulting in the creation of CodeT5, which is then fine-tuned on a vulnerability dataset. This approach improves upon VRepair's shortcomings in capturing diverse vulnerability features.In summary, current mainstream techniques focus on learning from richer and more diverse program data to acquire deeper vulnerability-related semantic information and use this knowledge to fine-tune on vulnerability datasets. However, these techniques often overlook the diversity of the vulnerability code itself. Specifically, they encode complex vulnerable code into a fixed vector through a simplified mapping, followed by direct decoding, which can lead to the loss of key vulnerability features and a lack of semantic depth.Additionally, for longer vulnerable code, these models rely on truncation, disregarding the precise location and contextual environment of the vulnerability, resulting in the loss of important information. This prevents effective capture of critical vulnerability features, limiting repair precision and significantly increasing the difficulty of searching for solutions during the repair process.Therefore, efficiently locating vulnerabilities and capturing diverse vulnerability features are crucial research areas for improving automatic vulnerability repair techniques.
\par
To address the limitations of existing automatic vulnerability repair techniques, this paper proposes CRepair, a multi-sample feature fusion technique based on a Conditional Variational Autoencoder (CVAE). Compared to VRepair and VulRepair, CRepair demonstrates higher accuracy and interpretability in both precise vulnerability localization and patch generation. This technique adopts the "chain of thought" approach\cite{7}, progressively repairing vulnerabilities by dividing the repair process into three distinct stages: data preprocessing to help the model quickly locate vulnerabilities, training to learn vulnerability features, and the final phase where the model generates repair patches.Specifically, during the data preprocessing stage, block-level vulnerabilities are serialized, and guiding prompt information is introduced to help the model swiftly locate and analyze the vulnerability type. In the training phase, the preprocessed data is fed into the model, and a multi-sample feature fusion technique is applied to enable the model to learn richer vulnerability feature information. This training strategy primarily aims to optimize model performance, improving both accuracy and generalization by minimizing the cross-entropy loss between the predicted and actual repair patches and reducing the KL divergence between the model’s predicted distribution of vulnerability features and the actual distribution.In the vulnerability repair stage, CRepair takes the vulnerable code as input and, using beam search, generates a specified number of candidate patches for cybersecurity analysts to review and select from. Various experiments have fully validated the effectiveness of the CRepair method, showcasing its broad potential for application in the field of automatic vulnerability repair.
\par
The main contributions of this paper are summarized as follows:

\begin{itemize}
	\item We propose CRepair, a CVAE-based automatic vulnerability repair technique designed to address cybersecurity vulnerabilities in development code, enabling automatic repair of vulnerable code.
	\item We introduce a probability distribution mapping technique that maps vulnerability feature data to probability distributions, addressing the limitations of existing methods that use fixed encoding, which leads to poor model generalization and insufficient ability to capture diverse vulnerability features.
	\item We propose a multi-sample feature fusion technique to overcome the challenge of insufficient vulnerability datasets, enabling the model to capture diverse vulnerability features even with limited data.
	\item We address the issues present in VRepair and VulRepair, and evaluate CRepair against mainstream models, achieving promising results.
\end{itemize}
\par
Section 1 of this paper introduces related work and the current state of research on vulnerability repair. Section 2 presents the necessary techniques and relevant background knowledge used in this study. Section 3 details the proposed CVAE-based multi-sample feature fusion model for automatic vulnerability repair, explaining the process of vulnerability learning and repair. Section 4 demonstrates the effectiveness of the proposed model through extensive experiments. Finally, Section 5 concludes the paper with a summary of the findings.

\section{RELATED WORK}
\par
Code vulnerabilities refer to serious flaws in the computational logic or code of a software system, which can be easily exploited to cause harm. Common examples include data breaches, system intrusions, and unauthorized modifications of privileges, all of which pose severe threats to system security. Timely repair of these vulnerabilities can prevent potential attacks, making vulnerability repair a core task in cybersecurity.To ensure system security, numerous researchers have conducted studies on vulnerability repair. For instance, Lee et al.\cite{8} proposed MemFix, an automated patch generation technique based on a two-step process of detection and repair. For each memory allocation statement, MemFix generates patches and combines them to repair vulnerabilities. However, this approach sacrifices efficiency and scalability, as it is limited to repairing specific types of vulnerabilities and results in a high false positive rate. To address this issue, researchers have proposed vulnerability repair based on dynamic analysis, where vulnerabilities are located and repaired while the code is being executed.
\par
Wang et al.\cite{9} addressed the limitations of static analysis in vulnerability detection by proposing a hybrid approach that combines both static and dynamic methods. They also employed taint analysis techniques to validate security vulnerabilities, thereby improving the efficiency of vulnerability remediation. However, the implementation of these methods is often complex and can lead to issues where the remediation does not meet expectations, making it difficult to ensure that vulnerabilities are properly fixed.As machine learning and deep learning technologies have matured, they have seen widespread application in fields such as image recognition, speech recognition, and natural language processing, achieving significant breakthroughs. In response to the challenges posed by traditional static and dynamic analysis-based vulnerability remediation—such as high false-positive rates and the considerable time and effort required from researchers—researchers have framed automatic vulnerability repair as a neural machine translation (NMT) task\cite{10}. The goal of NMT is to learn the mapping between vulnerable code and its correct, post-repair version. This is similar to the sequence-to-sequence (seq2seq) approach, which learns the mapping between two sequences\cite{11}, a method commonly used in neural machine translation, text summarization\cite{12}, and other natural language processing tasks.However, due to the limited availability of public vulnerability datasets and the large amounts of data required for training machine learning models, these methods struggle to learn effective information about vulnerabilities. This also makes it difficult to assess the effectiveness of such repair approaches.
\par
To address the scarcity of vulnerability datasets, Chen et al.[5] proposed VRepair, a vulnerability auto-repair technique based on transfer learning. VRepair collects vulnerability repair data by analyzing GitHub events, ultimately constructing a corpus of 650,000 C-language repair examples for pre-training. This model is then fine-tuned on CVEFixes and Big-Vul datasets, which alleviates the limitation of limited vulnerability data and demonstrates the effectiveness of transfer learning in addressing data scarcity. However, VRepair’s pre-training approach does not fully capture the diverse characteristics of vulnerabilities.To overcome this limitation, Mashhadi et al.\cite{13} introduced CodeBERT\cite{14}, a Transformer-based model with encoder and decoder structures (12 encoders and 6 decoders) designed for pre-training. It is pre-trained on CodeSearchNet—a large-scale corpus with 8.35 million functions across 8 programming languages\cite{15}—and later fine-tuned on vulnerability datasets. Though CodeBERT improves upon VRepair, its generalization capability remains limited.To tackle this issue, Fu et al.\cite{6} proposed VulRepair, a vulnerability auto-repair technique based on the T5 model. VulRepair applies four pre-training strategies on CodeSearchNet, resulting in CodeT5 as the foundational model, which is subsequently fine-tuned on vulnerability datasets. This approach addresses VRepair’s shortcomings in capturing diverse vulnerability features and utilizes the Byte-Pair Encoding (BPE) algorithm\cite{16} to handle rare vocabulary issues, boosting VRepair’s repair accuracy from 23\% to 44\%. This result highlights the immense potential of pre-trained models in vulnerability repair tasks.
\par
Despite the promising results achieved by the aforementioned techniques, they primarily extract a single feature from the vulnerable code. To address this limitation, they attempt to improve repair outcomes by learning from a larger quantity of vulnerability data, often overlooking the impact of the inherent diversity of vulnerabilities on repair effectiveness. Specifically, these techniques simplify the feature extraction process by directly mapping the vulnerable code into a fixed feature vector, which is then decoded for repair. However, relying on a single feature representation does not enable the model to fully learn from the vulnerable code, resulting in limited generalization capabilities.Moreover, when dealing with long vulnerabilities exceeding 512 tokens, these techniques typically process the vulnerability data homogeneously, applying truncation that can lead to the loss of critical vulnerability information. This compromises the model's ability to accurately identify and learn genuine vulnerability characteristics. Experimental analysis of VulRepair revealed that approximately 29\% of the dataset consists of long vulnerabilities, with 10\% of this data losing vulnerability information due to truncation, adversely affecting the model's repair performance.This paper aims to address the shortcomings of existing vulnerability repair techniques, which often neglect the diverse features of code and face challenges in accurately locating vulnerabilities in longer code snippets, leading to a larger search space. We propose corresponding improvements and optimizations, with the expectation that the model can learn diversely from limited data and quickly locate vulnerabilities in longer code, thereby enhancing both repair accuracy and generation quality. Next, we will provide a brief introduction to the relevant technologies involved in vulnerability repair.

\section{BACKGROUND KNOWLEDGE}
\par
The method proposed in this paper primarily employs Conditional Variational Autoencoders (CVAE) as the overall framework. It utilizes prompt engineering and the Byte-Pair Encoding (BPE) tokenization algorithm for preprocessing vulnerability data. The following sections will introduce the relevant concepts and foundational knowledge.
\subsection{Variational inference and conditional variational autoencoders}
\par
Conditional Variational Autoencoders (CVAE) are an extension of Variational Autoencoders (VAE)\cite{18}, designed as generative models that introduce conditional variables to precisely control specific features of the generated data. CVAE utilizes an encoder-decoder architecture, combined with sampling mechanisms for deep learning, which has led to its widespread application in fields such as image generation and natural language processing (NLP)\cite{19}. In the realm of text generation, CVAE is particularly significant as it not only enhances the model's ability to understand context but also improves the quality and generalization of the generated content. By employing conditional controls, CVAE ensures that the generated text meets specific conditions, resulting in more accurate and relevant outputs. Specifically, CVAE can implement variational inference through parameterized neural networks, learning the latent structure of data and constructing corresponding probability distributions. By sampling from these distributions, it captures the diverse features of the data, helping to mitigate issues of data scarcity and the singularity of feature information.
\par
In this paper, vulnerability data can be viewed as generated by unobservable latent variables that satisfy a certain specific distribution through random processes. This distribution represents the intrinsic features of the vulnerability code or the organizational structure between contexts. Let us denote the vulnerability code dataset as \(\{(x_i,y_i)|x_i\in X,y_i\in Y\}_{i=1}^{N}\),Here, X represents the vulnerability dataset, while Y denotes the corresponding repaired dataset. N is the total number of samples in the dataset. For the conditional variable c, this paper will extract feature information from the repaired code to guide the model in repairing the vulnerable code. Based on the aforementioned conditions, we employ variational inference to optimize the variational lower bound of the marginal likelihood, specifically the Evidence Lower Bound (ELBO), to enhance the model:
\[\mathcal{L} = \mathbb{E}_{q(z|x,c)}[\log{p(y|z,c)}]-D_{KL}(q(z|x,c)\|p(z|c))\]
\par
The first term represents the reconstruction error, which is the expected log-likelihood of the data generated by the decoder under the distribution of the latent variable z. Since the true posterior distribution \(p(z|x,c)\) is often intractable to compute directly, the second term utilizes variational inference to minimize the Kullback-Leibler (KL) divergence between the posterior distribution \(q(z|x,c)\) from the encoder and the prior distribution \(q(z|c)\), thereby driving the posterior distribution closer to the prior distribution. Here, \(q(z|x,c)\) indicates the distribution parameters of the latent variable z outputted by the encoder, which receives the input data x and the conditional variable c. This can be expressed using the following formula:
\[q(z|x,c) = \mathcal{N}(y;\mu ^{'}(z,c),\sigma^{'2}(z,c))\]
\par
\(\mathcal{N}\) indicates that a Gaussian distribution is used for calculations, where \(\mu ^{'}(z,c)\)  and \(\sigma^{'2}(z,c)\) represent the computed mean and variance, respectively. In this context, the distribution of z is an approximate probability distribution that aims to closely approximate the true posterior distribution by minimizing the Kullback-Leibler (KL) divergence with respect to the prior distribution \(p(z|c)\). This process helps construct a better representation of the distribution for the latent variable z, which is essential for subsequent sampling and decoding operations.
\par
In the first term, \(p(y|z,c)\) represents the distribution of the output y generated by the decoder, which receives the latent variable z and the conditional variable c. This distribution can also be computed using a Gaussian distribution, as expressed in the following formula:
\[p(y|z,c)=\mathcal{N}(y;\mu ^{'}(z,c),\sigma^{'2}(z,c))\]
\par
Here, \(\mu ^{'}(z,c)\) and \(\sigma^{'2}(z,c)\) are the mean and variance, respectively, computed by the decoder neural network. Based on the above, the generation process in CVAE consists of the following two steps:
\begin{enumerate}[(1)]
	\item The probability distribution \(q(y|z,c)\) of the latent variable z is constructed using the condition c and the data x, and z is sampled from this distribution.
	\item The decoder \(p(y|z,c)\), conditioned on c, generates the output y based on the latent variable z.
\end{enumerate}
\par
The core of CVAE lies in learning a latent variable representation z by combining the conditional variable c with the input data x, enabling the generation of conditioned data y. During training, by continuously optimizing the variational lower bound, the robustness and diversity of the learned latent features are significantly improved. This allows for precise control over the generated data, ultimately enhancing the quality of the generated samples.

\subsection{Prompt Engineering}
\par
Prompt engineering is a technical approach that utilizes prompts to facilitate knowledge transfer during task execution, similar to a software pattern. This method enhances the model's understanding of tasks by providing rich feature information specific to the context\cite{17}. During model training, prompts can be added to guide the model to learn particular tasks or behaviors. As shown in Figure 2-c, this paper introduces prompt engineering during the data preprocessing phase. Specifically, CWE vulnerability prompt information is embedded at the beginning of the vulnerable source code. This vulnerability type prompt helps the model quickly identify the type of vulnerability and develop a corresponding repair strategy. Special prompt markers such as \textless StartLoc \textgreater and \textless EndLoc \textgreater are then added around the start and end locations of the vulnerability, enabling the model to more accurately pinpoint the vulnerability and focus on learning its features.By applying prompt engineering, the CRepair model can tailor its learning to different types of vulnerabilities, and quickly and precisely locate vulnerabilities using prompts. This enhances the accuracy of feature extraction and improves the effectiveness of vulnerability repair.

\subsection{BPE Algorithm}
\par
Byte Pair Encoding (BPE) is an unsupervised subword tokenization technique widely used in neural machine translation and natural language processing tasks\cite{20}. This technique reduces the vocabulary size by iteratively merging the most frequent byte pairs in the text, effectively addressing the issue of out-of-vocabulary (OOV) words. Specifically, BPE operates in two key steps: character pair merging and vocabulary construction. First, the algorithm breaks down longer words, selects and merges the most frequent character pairs to form new words. These newly formed words are then added to the vocabulary. For instance, when tokenizing an unfamiliar function name like "calculate\_total," the output might be the list [“calculate”, “total”]. By reorganizing common subwords, BPE reduces vocabulary size, enhancing the model's generalization and flexibility. In this paper, since code text often contains a large number of repetitive structures and patterns, it is highly likely to encounter new identifiers or variable names that are not in the vocabulary. Therefore, BPE is employed to process the code, splitting unfamiliar words into smaller subunits, thereby reducing sparsity and making the model's training process more efficient while improving the generalization of the learned representations.

\section{CREPAIR: CVAE-BASED AUTOMATIC VULNERABILITY REPAIR TECHNOLOGY}
\subsection{Problem Definition}
\par
The goal of this research is to develop a model that can automatically repair vulnerable code by extracting and analyzing the intrinsic features of vulnerability code diversity. Specifically, the vulnerable code (highlighted in red in Figure 2-a) is preprocessed and used as input to the model, with the expectation that the model will output the repaired code, as shown in Figure 2-d. This output is then reviewed by security analysts, who can further refine the vulnerable code into the correct version, as illustrated in Figure 2-b, with the green parts representing the repaired content. In this paper, the vulnerability code dataset is represented as \(\{(x_i,y_i)|x_i\in X,y_i\in Y\}_{i=1}^{N}\), where X denotes the set of vulnerable code, Y represents the corresponding set of repaired code, and N is the total number of samples in the dataset.

\subsection{Model Building}
\par
This paper proposes a vulnerability auto-repair technique, CRepair, based on Conditional Variational Autoencoders (CVAE), aimed at addressing the limitations of existing methods that overlook the impact of intrinsic diversity in vulnerability code on repair effectiveness, as well as the issue of large search space caused by slow vulnerability localization. The overall model architecture and repair process are shown in Figure 1. Specifically, the vulnerability repair process is divided into three stages: data preprocessing, model training, and vulnerability repair.
\par
In the data preprocessing stage, code block-level vulnerabilities are first expanded into sequences, followed by embedding prompt markers. BPE (Byte Pair Encoding) is then applied to tokenize the code, which serves as input to the model. During the model training phase, CRepair utilizes the T5 encoder to encode the data and extract vulnerability features. It then computes the mean and variance of these features to construct a probability distribution of the vulnerability characteristics. By sampling from this distribution multiple times, different vulnerability feature samples are obtained, which are then fused to generate the latent variable z. This latent variable, combined with the conditional variable c, is passed to the T5 decoder for decoding. The conditional variable c is derived by extracting features from the true repair samples y. After decoding, linear expansion and maximum probability calculations are used to obtain potential repair patches. The primary goal of the training process is to deeply learn the diverse features of vulnerability code, enabling precise control over the generated repairs through the conditional variable. The model is optimized by maximizing the variational lower bound of the marginal likelihood, which reduces both cross-entropy and KL divergence. In the vulnerability repair phase, the trained CRepair model receives the vulnerable code as input and outputs a ranked set of repair patches through beam search, which serves as a candidate repair set for security analysts to evaluate and implement.The following sections provide a detailed explanation of these three stages.
\begin{figure}
	\centering
	\includegraphics[width=1\linewidth]{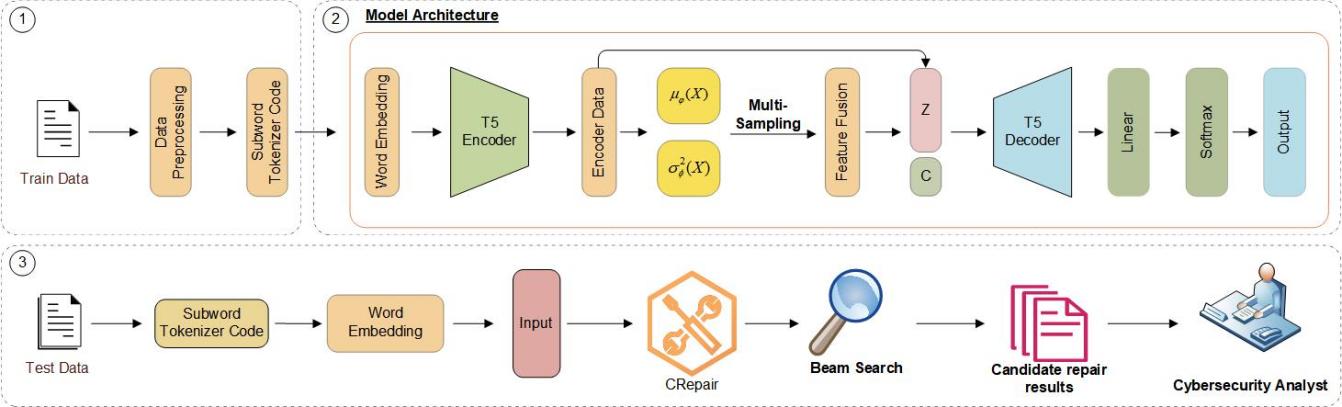}
	\caption{CRepair model architecture and workflow}
	\label{fig:enter-label}
\end{figure}

\subsection{Data Preprocessing}
\par
The data preprocessing stage is primarily aimed at reducing noise and outliers in the data, preparing it into a format that the model can accept and use. This ensures that the model can effectively learn from the data, enhancing its generalization capability, performance, and practical utility. The preprocessing in this paper focuses on the following aspects:
\par
\textbf{Code Serialization:} By converting code blocks into a sequence, it reduces the overhead of control loops and reveals the complete structure and logic of the code. This allows the model to better understand the code's context and execution flow, improving its comprehension of code semantics. Specifically, the original block-level vulnerability (as shown in Figure 2-a) is transformed into a continuous text sequence, removing comments and other distracting noise so that the model can focus on learning the key vulnerability features.
\par
\textbf{Prompt Insertion:} Adding prompt information to the code sequence helps the model quickly identify the critical features and intent of the code, guiding it to focus on the most important parts of the code. This accelerates the localization of vulnerabilities and prioritizes their learning. Specifically, vulnerabilities are first classified according to their type, and a corresponding CWE vulnerability category is inserted at the beginning of the code sequence as a type prompt. Then, markers indicating the vulnerability’s start and end, such as \textless StartLoc \textgreater and \textless EndLoc \textgreater (as shown in Figure 2-c), are embedded to help the model locate the vulnerability. This approach allows the model to tailor different repair strategies for various types of vulnerabilities and quickly focus on them, improving the precision of vulnerability feature extraction.

\par
\textbf{Tokenization:} Given that the syntax structure of code is significantly different from regular text, traditional tokenization methods often mark complex or custom long words as unknown [UNK]. By applying Byte Pair Encoding (BPE), the code text is split into smaller subword units, which addresses the out-of-vocabulary (OOV) issue. This helps the model capture both the semantic and syntactic features of the code, enhancing the accuracy of feature extraction. In this study, the vulnerability code sequences are tokenized using BPE, converting the code into continuous vector representations as input to the model. This tokenization reduces the vocabulary size that the model needs to learn, easing the model's burden and improving data processing efficiency, which in turn strengthens the model's generation and comprehension capabilities.

\begin{figure}
	\centering
	\includegraphics[width=1\linewidth]{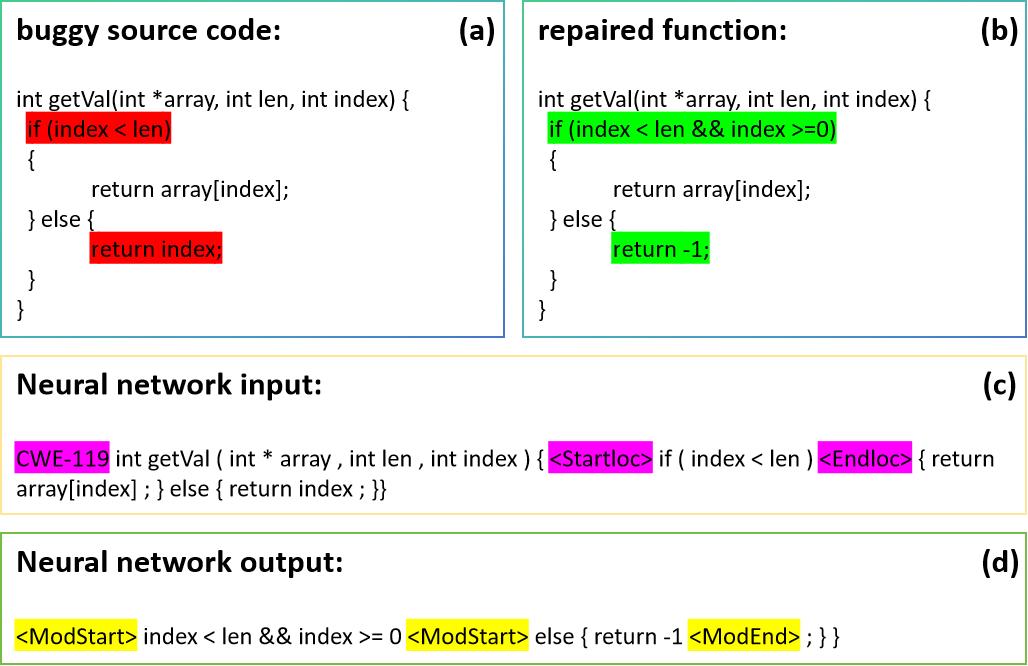}
	\caption{Enter Caption}
	\label{fig:enter-label}
\end{figure}

\subsection{Model Training}
\par
In the model training phase, CRepair is built on the Conditional Variational Autoencoder (CVAE) framework, utilizing the encoder and decoder modules from the T5 model. Specifically, CRepair first uses the encoder to extract vulnerability features, including the mean and variance of the vulnerability distribution, which are then used to construct a probabilistic distribution of the vulnerability characteristics. Next, by sampling from this distribution, the model captures the feature information of the vulnerability and uses this as the input vector for the decoder. Finally, the decoder processes these features to generate repair patches. In this work, a multi-sample feature fusion technique is introduced during the sampling process, where multiple samples from the probabilistic distribution of vulnerability features are merged, enhancing the diversity of the captured features. Through these training steps, the goal is for the model to output repair patches as shown in Figure 2-d. Additionally, the model employs cross-entropy loss\cite{21} and KL divergence to guide the gradient descent process, further optimizing the model's performance. The following sections will provide a detailed explanation of the CRepair implementation.
\par
\textbf{Encoder.} In this work, the encoder is primarily used to encode vulnerability code, enabling extraction of the global features of vulnerabilities and generating a high-dimensional feature embedding representation. Since Transformers\cite{22} can handle the contextual information of input code, taking into account overall code structure and dependencies, the encoder comprises 12 stacked Transformer blocks with identical structures. This encoder is designed to process tokenized subword representations from BPE (notated as \(Z = \{Z_1,Z_2,...,Z_n\}\)) to capture the global features of the vulnerability. Recognizing the importance of each token’s dependency and positional information within code logic, relative positional encoding is incorporated in the encoding process, allowing for more precise capture of contextual and positional relationships between tokens. The resulting encoding is a highly aggregated representation encapsulating the key feature information of the original data. The Transformer's encoder structure includes a multi-head self-attention layer\cite{23} and a feed-forward neural network, with layer normalization\cite{24} to standardize data processing. In the self-attention mechanism, a dot-product operation calculates attention scores for each token, enabling interaction between a token and every other token in the sequence. This mechanism relies on three vectors: Query, Key, and Value,where Query represents the focus of the current token on other positions, Key is the matching vector for score calculation, generating attention weights, and Value combines with attention weights to update token representations. Additionally, relative positional encoding is integrated into self-attention to account for token distances, enhancing the model’s ability to understand code structure. The calculation process is as follows, where Q, K, V represent Query, Key, and Value, respectively,  is the encoding dimension, and P is the positional encoding.

\[Attention(Q,K,V) = softmax(\frac{Q(K+P)^T}{\sqrt{d_k}}(V+P) )\]

\par
To more effectively capture the semantic information within sequences, this study employs a multi-head self-attention mechanism to process the data, allowing the model to consider vulnerability information from multiple perspectives. The d-dimensional Query (Q), Key (K), and Value (V) vectors are each divided into h heads, with each head maintaining  dimensions. After each head independently performs self-attention operations, the results are concatenated and passed to a feed-forward neural network for further processing. This approach enables the model to learn rich feature representations in different subspaces, thereby improving the accuracy and efficiency of information extraction. The calculation formulas are as follows, where  represents the weight matrix for the linear transformation of the output.

\[MultiHead(Q,K,V) = Concat(head_1,...,head_n)W^O\]

\par
Finally, a Feed-Forward Network (FFN) is used to process the input data, allowing the model to learn more complex features through the introduction of nonlinear transformations, thereby enhancing its learning capacity. The output from each of the 12 layers of the Transformer encoder serves as the input for the subsequent layer. At the end of this process, the output from the final layer is extracted as the hidden state, referred to as EncoderData.

\par
\textbf{Multi-sampling feature fusion.} Traditional encoder structures typically map data directly to a fixed feature vector, which is then used for decoding. However, this approach can lead to a lack of flexibility and an increased risk of overfitting, especially in small-sample learning scenarios. To address this issue, this study maps vulnerability features onto a probability distribution. Specifically, it first employs an adaptive weighted attention mechanism to extract the mean  and variance  of the vulnerability features, thereby constructing the corresponding probability distribution for sampling. The process is illustrated in Figure 3.After obtaining the hidden state, referred to as EncoderData, from multiple layers of feature extraction by the encoder, this data is input into the adaptive weighted attention module for processing. This module initially computes the dot product between the hidden state and the attention weight tensor to generate scores for each position. Subsequently, these scores are masked to ignore irrelevant confounding elements, retaining only the key features associated with the vulnerabilities, which facilitates focused learning of their local characteristics. The Softmax function is then applied to these scores to obtain the attention weights, and finally, a weighted average of the hidden states and the attention weights is computed to yield the weighted data. This approach enriches the contextual information of the data.Since convolutional operations are more effective at capturing local vulnerability features, the processed weighted data undergoes convolution to derive the mean  and variance of the vulnerability distribution, thereby forming a high-quality probability distribution for the vulnerability features. To address the non-differentiability issue in variational inference, this study employs the reparameterization\cite{25} technique to sample potential variable Z based on the mean and variance, computed as \(Z = \mu + \sigma * \epsilon \), where x follows a standard normal distribution. However, single sampling is insufficient to capture a sufficiently diverse set of vulnerability features. Therefore, this study utilizes multiple sampling and multi-angle analysis to obtain richer local feature information related to vulnerabilities. Subsequently, the mean fusion calculation is performed on these feature sets, which can be expressed as:
\[Z = \frac{\sum_{i=0}^{n}Z_1 }{n} \]

\par
Here, n represents the number of samples. By performing multiple samplings, the sampled data can closely approximate the true distribution, enhancing the model's robustness. This approach also enables data augmentation in scenarios with limited data, thereby improving the quality of decoding and alleviating the overfitting issues often associated with small datasets. To facilitate the model's ability to learn vulnerability feature information more effectively, this study computes a residual between the fused latent variable Z and the encoded EncoderData. This method, which combines the globally encoded features with diverse local features, effectively integrates the original input and the encoded features, allowing for the retention of more vulnerability feature information.

\par
\textbf{Conditional control.} In a Conditional Variational Autoencoder (CVAE), conditional variables are used to guide the model's generation process, enabling it to produce samples based on specific conditions. This study primarily extracts features from real repair samples y to obtain the conditional variable c, which directs the model in the vulnerability repair process. As shown in Figure 3, we created a feature extraction module called Extract Features to process the real repair samples y. Specifically, this module consists of a positional encoding layer, a data encoding layer, and a linear activation layer. The positional encoding layer employs relative positional encoding to help the model better learn the contextual relationships between tokens. The data encoding layer encodes the repair data, producing repair feature data of the same dimension as the latent variable. Subsequently, a linear activation function is applied to the data, resulting in the conditional variable c, which contains information about the repaired vulnerabilities. This process provides additional context for subsequent decoding, ensuring that the generated data aligns with specific conditions and allowing for greater flexibility and control in the model's generation operations.

\begin{figure}
	\centering
	\includegraphics[width=0.5\linewidth]{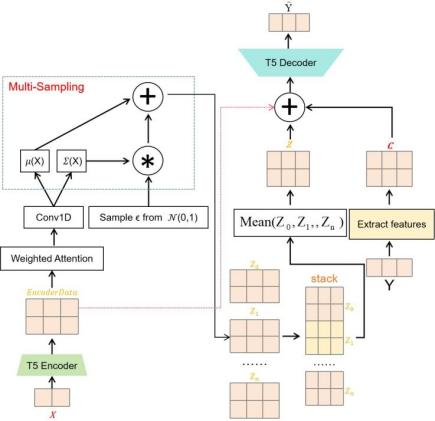}
	\caption{Enter Caption}
	\label{fig:enter-label}
\end{figure}

\par
\textbf{Decoder.} The primary function of the decoder is to decode the sampled vulnerability feature data to generate vulnerability repair patches. Its structure is similar to that of the encoder, consisting of 12 stacked Transformer blocks. Each decoder block is composed of four components: a normalization layer, a masked multi-head self-attention layer, a multi-head self-attention layer, and a feed-forward neural network.In this study, the sampled latent variable Z, the encoded data from the encoder (referred to as Encoder Data), and the conditional variable C are combined as inputs to the first layer of the decoder. This approach effectively integrates the global and local features of vulnerabilities, with the global features providing overall context and the local features focusing on specific details of the vulnerabilities. As a result, the decoder is able to capture a richer set of vulnerability feature information. Additionally, the introduction of the conditional variable effectively guides the decoder’s repair process. The specific calculation process is as follows, where E represents the encoded data, Z denotes the sampled data, and C signifies the conditional variable.
\[O_1=FFN(MultiHead(Z_1))\]
\[Z_1=MaskMultiHead(LayerNorm(E,Z,C))\]

\par
Masked multi-head self-attention ensures that the model does not access future information (i.e., subsequent tokens) when predicting the current token. This allows the model to generate tokens step by step, enabling the automated repair of vulnerability code layer by layer. Once the first decoder block completes its decoding, its output is used as the input for the next layer, continuing the generation process. The specific calculation process is outlined as follows.
\[O_1=FFN(MultiHead(Z_i))\]
\[Z_i=MaskMultiHead(LayerNorm(E,Z_{i-1},C))\]

\par
Once the final decoder block has completed its decoding, the last output is linearly transformed to expand the embedding dimensions of each decoded token to the size of the vocabulary. This is followed by a softmax layer, which calculates the probabilities of the decoded results, enabling the model to generate its predictions.

\par
\textbf{Loss calculation.} The loss function for CRepair consists mainly of reconstruction loss (RC) and KL divergence loss (KL). The reconstruction loss measures the cross-entropy between the model's output and the target data, assessing the accuracy of the generated data. In contrast, the KL loss evaluates the difference between the distribution of the generated latent variables Z and the prior distribution of the real vulnerability data. The goal is to make the sampled data closer to the actual input data, ensuring that vulnerability features are preserved, which facilitates learning diverse vulnerability characteristics. The specific calculation of the loss is as follows.

\[\mathcal{L}_{CVAE}=\mathcal{L}_{RC}+\mathcal{L}_{KL}\]

\subsection{vulnerability repair}
\par
After the model training is complete, this paper employs a beam search strategy\cite{26} during the generation phase. Specifically, beam search initially selects multiple candidates, and at each time step, it retains the top few tokens with the highest probabilities as candidates. Based on the set beam width, only the highest-probability candidates are kept while others are discarded. After multiple iterations, the final candidate repair set is selected based on probability rankings. Finally, the decoded results are tokenized and provided to security analysts for review and selection.

\section{EXPERIMENTAL}
\subsection{Dataset Introduction And Experimental Settings}
\par
This section presents the experimental results of CRepair. CRepair is compared against baseline models such as VRepair, CodeBERT, and VulRepair, as these methods and models are open-source and represent current mainstream techniques for automated vulnerability repair. The vulnerability datasets used include CVEFixes\cite{27} and Big-Vul\cite{28}. By combining these two datasets, a total of 6,844 training samples and 1,638 test samples were obtained. In the experiments, the training data was divided into a training set and a validation set, with 12.38\% of the data allocated for validation and the remainder for training. The deep learning framework for this experiment is PyTorch, and the training and testing were conducted on an NVIDIA GeForce RTX 3090. The main parameters of CRepair are summarized in Table 1.

\begin{table}
	\centering
	\begin{tabular}{cc}
		\hline
		Parameter & Setting\\
		\hline
		Encoding Dim & 512\\
		Decoding Dim & 256\\
		Epochs & 75\\
		Learning Rate & 2e-5\\
		Optimizer & Adam\\
		Batch Size & 8\\
		Beam Num & 50\\
		Embedding Dim & 768\\
		Sample Size & 5\\
		\hline
	\end{tabular}
	\caption{Main parameter settings of the model}
	\label{tab:my_label}
\end{table}

\subsection{Evaluation indicators}
\par
To facilitate comparison with baseline models, this section establishes a unified evaluation metric for the models. This study employs the "Perfect Repair Percentage" as the criterion for assessing the accuracy of the models' repairs. A repair is considered successful, or a perfect prediction, only when the candidate repair set output by the model contains content that is completely identical to the target repair patch. To evaluate the model's repair effectiveness, this study tests and assesses 1,638 vulnerability samples. The specific evaluation method involves calculating the ratio of the number of vulnerabilities with perfect repairs to the total number of vulnerabilities in the test dataset. The calculation method is as follows:

\[Perfect \: Repair\:Ratio = \frac{Number \: of \:perfectly \: fixed \: vulnerabilities}{Total \: number \: of \: vulnerabilities} \]

\par
In this study, a higher value of the Perfect Repair Percentage indicates better repair performance of the model.

\subsection{Problem exploration and experimentation}
\par
To evaluate the effectiveness of the CVAE-based multi-sampling feature fusion vulnerability automatic repair technique, CRepair, this chapter will discuss the following questions:
\begin{itemize}
	\item  RQ1: Does CRepair achieve better results in vulnerability repair compared to baseline models?
	\item RQ2: What is the contribution of each component of the CRepair model?
	\item RQ3: What is the optimal number of sampling iterations for CRepair?
\end{itemize}
\par
\textbf{RQ1: Does CRepair achieve better results in vulnerability repair compared to baseline models?}
\par
To investigate the differences in vulnerability repair effectiveness between CRepair and other baseline models, and to address the first research question posed in this paper, we will conduct a comparative experiment on the repair accuracy of CRepair against VRepair, CodeBert, and VulRepair. These experiments aim to provide a comprehensive assessment of the models' repair capabilities. To ensure fairness, all models will be evaluated using the same experimental setup and datasets, with the beam search width set to 50 during text generation. Table 2 presents the comparative experimental results of CRepair and the other baseline models regarding repair effectiveness.

\begin{table}
	\centering
	\begin{tabular}{cc}
		\hline
		Model & Accuracy\\
		\hline
		VRepair & 0.2321\\
		CodeBert & 0.3142\\
		VulRepair & 0.4408\\
		\textbf{\underline{CRepair}} & \textbf{\underline{0.5189}}\\
		\hline
	\end{tabular}
	\caption{Comparison experiment of repair effects between CRepair and other benchmark models}
	\label{tab:my_label}
\end{table}
\par
Experimental analysis shows that CRepair achieves the best repair performance, successfully fixing 850 out of 1,638 vulnerabilities, resulting in a repair accuracy of 52\%. This demonstrates that CRepair outperforms the baseline models in terms of vulnerability repair effectiveness.

\par
\textbf{RQ2: What is the contribution of each component of the CRepair model?}
\par
The prompt method, conditional control, and multi-sampling feature fusion techniques are core components of this study. To investigate the different impacts of these methods on the model, this section will conduct ablation experiments comparing CRepair with other baseline models. Specifically, four types of experiments will be performed: (1) using the prompt method without sampling; (2) not using the prompt method and without sampling; (3) using multi-sampling with the prompt method; and (4) using multi-sampling without the prompt method. Since the multi-sampling feature fusion method is an implementable part of the CVAE architecture, only CRepair will undergo ablation experiments, while the baseline models will not use this method. To verify the effect of conditional control, a comparative experiment will be conducted between VAE-Repair, which does not use the conditional variable, and CRepair. The experimental results are shown in Table 3.
\par

\begin{table}
	\centering
	\begin{tabular}{cccccc}
		\hline
		Method & CRepair & VAE-Repair & VRepiar & CodeBert & VulRepair\\
		\hline
		NP & 0.4621 & 0.4548 & 0.2321 & 0.2011 & 0.4408\\
		NN & 0.3834 & 0.3742 & 0.1623 & 0.1710 & 0.3624\\
		\textbf{\underline{MP}} & \textbf{\underline{0.5189}} & 0.4774 & / & / & / \\
		MN & 0.4304 & 0.4191 & / & / & / \\
		\hline
	\end{tabular}
	\caption{Module ablation experiment. In this table, NN represents the combination of no sampling and no prompting, while NN represents the combination of no sampling and no prompting, MP represents the combination of multiple sampling and prompting, and MN represents the combination of multiple sampling and no prompting.}
	\label{tab:my_label}
\end{table}

\par
The experimental data indicate that when the model does not perform sampling, the repair performance of CRepair, VAE-Repair, VRepair, CodeBERT, and VulRepair improves by 21\%, 22\%, 43\%, 17\%, and 21\%, respectively, compared to the case without prompts. Notably, VRepair exhibits the most significant improvement, followed closely by CRepair, while CodeBERT shows the least enhancement. We speculate that this is because VRepair's pre-training approach fails to adequately capture the diverse features of vulnerabilities, and the addition of prompts helps the model quickly locate vulnerabilities, thereby reducing the search space for repairs and better capturing vulnerability features. This effect is also reflected in CRepair, likely due to the CVAE structure's strong reliance on prompts. Providing clear and high-quality prompts aids the model in generating more accurate repair data. The experimental results demonstrate that the prompt method significantly enhances the model's understanding of vulnerabilities, enabling it to quickly locate them and learn more vulnerability feature information.
\par
To validate the guiding role of the conditional variable in the model's repair generation process, we trained a VAE-Repair model without conditional control and compared it with CRepair. The experimental results indicate that when the conditional variable is introduced, CRepair significantly outperforms VAE-Repair in all aspects. Furthermore, when both multi-sampling and prompt addition are utilized, CRepair shows an 8\% improvement in repair effectiveness. This demonstrates that the conditional variable effectively guides the model in repairing vulnerabilities, enhancing the capabilities of traditional VAE to handle more complex generation tasks.
\par
To validate the effectiveness of the proposed multi-sampling feature fusion technique, we conducted multiple experiments with CRepair. The results indicate that when using prompts, the application of multi-sampling feature fusion improves CRepair's performance by 12\%. In the absence of prompts, the performance increase reaches 15\%. This outcome demonstrates the effectiveness of this technique, achieving significant improvements even with limited data. Overall, the results of the ablation experiments show that when the prompt method, conditional control, and multi-sampling feature fusion techniques are applied simultaneously, the model's repair effectiveness is optimized, with an enhancement of up to 39\%. This finding further confirms the effectiveness of the techniques proposed in this study, significantly improving vulnerability repair outcomes. To further explore the performance of the multi-sampling feature fusion technique at different sampling frequencies, we will conduct a hyperparameter analysis in the next section to determine the optimal sampling count.

\par
In summary, the method proposed in this paper effectively addresses the issues of fixed encoding overlooking the semantic diversity of code and the excessive search space caused by the inability to quickly locate vulnerabilities, particularly in scenarios where vulnerability datasets are limited. This approach represents a significant contribution to the field of automated vulnerability repair, with the potential to be applied for the identification and remediation of a broader range of vulnerabilities.

\par
\textbf{RQ3: What is the optimal number of sampling iterations for CRepair?}
\par
Since varying sampling quantities can have different impacts on model performance, this paper conducts multiple experiments with different sampling counts to determine the optimal sampling value and calculates the average perfect repair percentage. Table 4 presents the results of the vulnerability repair experiments conducted with different sampling quantities.

\begin{table}
	\centering
	\begin{tabular}{cccccc}
		\hline
		Sample Size & 1 & 3 & 5 & 7 & 9\\
		\hline
		Average Accuracy & 0.4634 & 0.4716 & 0.5086 & 0.4825 & 0.4847\\
		\hline
	\end{tabular}
	\caption{ Comparison of the effects of different sampling times on vulnerability repair}
	\label{tab:my_label}
\end{table}
\par
The experimental results indicate that when the sampling quantity is too low, particularly at a sampling count of one, the model fails to acquire sufficient diversity in vulnerability feature information, resulting in the poorest repair performance. The best repair outcomes are observed with five samples, where each experiment achieves approximately a 50\% repair rate. However, this does not imply that a higher sampling count always yields better results. As the number of samples increases, repair performance declines, likely due to the introduction of excessive noise that diminishes the significance of vulnerability feature information, thereby affecting the model's ability to learn these features and increasing uncertainty and variability during the training process.
\par
In summary, the experimental results indicate that the model achieves optimal performance with five sampling iterations. This validates the rationale and interpretability of the sampling count and demonstrates that prompt engineering and multi-sampling feature fusion techniques effectively assist the model in quickly identifying vulnerabilities and extracting diverse vulnerability features, thereby enhancing overall vulnerability repair performance.

\section{CONCLUSION}
\par
Automated vulnerability repair aims to help developers effectively address vulnerabilities in systems, reducing labor costs and achieving automatic fixes, thereby enhancing data security and mitigating the risks associated with cyber intrusions, which can lead to severe losses. This paper presents CRepair, an automated vulnerability repair technique based on Conditional Variational Autoencoders (CVAE), designed to tackle the issues in existing repair technologies that neglect code semantic diversity due to fixed encoding and struggle with large search spaces resulting from the inability to quickly locate vulnerabilities.Utilizing the CVAE architecture, we preprocess vulnerable code using prompts as model inputs and employ multi-sampling feature fusion techniques to diversify the extraction of vulnerability features. Concurrently, we utilize conditional control to guide the model in accurately repairing vulnerabilities. The generated candidate repair sets provide developers with faster and more precise support for vulnerability fixes. Through extensive experimentation, this study demonstrates that CRepair not only effectively repairs vulnerabilities but also significantly improves upon current mainstream repair technologies, showcasing promising application prospects.

\section*{Acknowledgments}
\par
This work is supported by the National Natural Science Foundation of China under grant number 62067007.

\bibliographystyle{elsarticle-num} 
\bibliography{ref.bib}

\begin{thebibliography}{10}
\expandafter\ifx\csname url\endcsname\relax
  \def\url#1{\texttt{#1}}\fi
\expandafter\ifx\csname urlprefix\endcsname\relax\def\urlprefix{URL }\fi
\expandafter\ifx\csname href\endcsname\relax
  \def\href#1#2{#2} \def\path#1{#1}\fi

\bibitem{1}
M.~Dowd, J.~McDonald, J.~Schuh, The art of software security assessment:
  Identifying and preventing software vulnerabilities, Pearson Education, 2006.

\bibitem{2}
W.~Hu, V.~L. Thing, Cpe-identifier: Automated cpe identification and cve
  summaries annotation with deep learning and nlp, arXiv preprint
  arXiv:2405.13568 (2024).

\bibitem{3}
S.~Christey, J.~Kenderdine, J.~Mazella, B.~Miles, Common weakness enumeration,
  Mitre Corporation (2013).

\bibitem{4}
Y.~Nong, Y.~Ou, M.~Pradel, F.~Chen, H.~Cai, Generating realistic
  vulnerabilities via neural code editing: an empirical study, in: Proceedings
  of the 30th ACM Joint European Software Engineering Conference and Symposium
  on the Foundations of Software Engineering, 2022, pp. 1097--1109.

\bibitem{5}
Z.~Chen, S.~Kommrusch, M.~Monperrus, Neural transfer learning for repairing
  security vulnerabilities in c code, IEEE Transactions on Software Engineering
  49~(1) (2022) 147--165.

\bibitem{6}
M.~Fu, C.~Tantithamthavorn, T.~Le, V.~Nguyen, D.~Phung, Vulrepair: a t5-based
  automated software vulnerability repair, in: Proceedings of the 30th ACM
  joint european software engineering conference and symposium on the
  foundations of software engineering, 2022, pp. 935--947.

\bibitem{7}
P.~Lu, S.~Mishra, T.~Xia, L.~Qiu, K.-W. Chang, S.-C. Zhu, O.~Tafjord, P.~Clark,
  A.~Kalyan, Learn to explain: Multimodal reasoning via thought chains for
  science question answering, Advances in Neural Information Processing Systems
  35 (2022) 2507--2521.

\bibitem{8}
J.~Lee, S.~Hong, H.~Oh, Memfix: static analysis-based repair of memory
  deallocation errors for c, in: Proceedings of the 2018 26th ACM Joint meeting
  on European software engineering conference and symposium on the foundations
  of software engineering, 2018, pp. 95--106.

\bibitem{9}
W.~Chao, L.~Qun, W.~XiaoHu, R.~TianYu, D.~JiaHan, G.~GuangXin, S.~EnJie, An
  android application vulnerability mining method based on static and dynamic
  analysis, in: 2020 IEEE 5th Information Technology and Mechatronics
  Engineering Conference (ITOEC), IEEE, 2020, pp. 599--603.

\bibitem{10}
Y.~Wu, M.~Schuster, Z.~Chen, Q.~V. Le, M.~Norouzi, W.~Macherey, M.~Krikun,
  Y.~Cao, Q.~Gao, K.~Macherey, et~al., Google's neural machine translation
  system: Bridging the gap between human and machine translation, arXiv
  preprint arXiv:1609.08144 (2016).

\bibitem{11}
R.~Nallapati, B.~Zhou, C.~Gulcehre, B.~Xiang, et~al., Abstractive text
  summarization using sequence-to-sequence rnns and beyond, arXiv preprint
  arXiv:1602.06023 (2016).

\bibitem{12}
R.~Nallapati, B.~Zhou, C.~Gulcehre, B.~Xiang, et~al., Abstractive text
  summarization using sequence-to-sequence rnns and beyond, arXiv preprint
  arXiv:1602.06023 (2016).

\bibitem{13}
E.~Mashhadi, H.~Hemmati, Applying codebert for automated program repair of java
  simple bugs, in: 2021 IEEE/ACM 18th International Conference on Mining
  Software Repositories (MSR), IEEE, 2021, pp. 505--509.

\bibitem{14}
Z.~Feng, D.~Guo, D.~Tang, N.~Duan, X.~Feng, M.~Gong, L.~Shou, B.~Qin, T.~Liu,
  D.~Jiang, et~al., Codebert: A pre-trained model for programming and natural
  languages, arXiv preprint arXiv:2002.08155 (2020).

\bibitem{15}
H.~Husain, H.-H. Wu, T.~Gazit, M.~Allamanis, M.~Brockschmidt, Codesearchnet
  challenge: Evaluating the state of semantic code search, arXiv preprint
  arXiv:1909.09436 (2019).

\bibitem{16}
R.~Sennrich, Neural machine translation of rare words with subword units, arXiv
  preprint arXiv:1508.07909 (2015).

\bibitem{18}
J.~Gao, W.~Bi, X.~Liu, J.~Li, G.~Zhou, S.~Shi, A discrete cvae for response
  generation on short-text conversation, arXiv preprint arXiv:1911.09845
  (2019).

\bibitem{19}
X.~Shen, H.~Su, S.~Niu, V.~Demberg, Improving variational encoder-decoders in
  dialogue generation, in: Proceedings of the AAAI conference on artificial
  intelligence, Vol.~32, 2018.

\bibitem{17}
J.~White, Q.~Fu, S.~Hays, M.~Sandborn, C.~Olea, H.~Gilbert, A.~Elnashar,
  J.~Spencer-Smith, D.~C. Schmidt, A prompt pattern catalog to enhance prompt
  engineering with chatgpt, arXiv preprint arXiv:2302.11382 (2023).

\bibitem{20}
M.~Gall{\'e}, Investigating the effectiveness of bpe: The power of shorter
  sequences, in: Proceedings of the 2019 conference on empirical methods in
  natural language processing and the 9th international joint conference on
  natural language processing (EMNLP-IJCNLP), 2019, pp. 1375--1381.

\bibitem{21}
L.~Li, M.~Doroslova{\v{c}}ki, M.~H. Loew, Approximating the gradient of
  cross-entropy loss function, IEEE access 8 (2020) 111626--111635.

\bibitem{22}
H.~Yan, B.~Deng, X.~Li, X.~Qiu, Tener: adapting transformer encoder for named
  entity recognition, arXiv preprint arXiv:1911.04474 (2019).

\bibitem{23}
E.~Voita, D.~Talbot, F.~Moiseev, R.~Sennrich, I.~Titov, Analyzing multi-head
  self-attention: Specialized heads do the heavy lifting, the rest can be
  pruned, arXiv preprint arXiv:1905.09418 (2019).

\bibitem{24}
J.~L. Ba, Layer normalization, arXiv preprint arXiv:1607.06450 (2016).

\bibitem{25}
R.~Tian, Y.~Mao, R.~Zhang, Learning vae-lda models with rounded
  reparameterization trick, in: Proceedings of the 2020 Conference on Empirical
  Methods in Natural Language Processing (EMNLP), 2020, pp. 1315--1325.

\bibitem{26}
Y.~Zhou, C.~Cui, J.~Yoon, L.~Zhang, Z.~Deng, C.~Finn, M.~Bansal, H.~Yao,
  Analyzing and mitigating object hallucination in large vision-language
  models, arXiv preprint arXiv:2310.00754 (2023).

\bibitem{27}
G.~Bhandari, A.~Naseer, L.~Moonen, Cvefixes: automated collection of
  vulnerabilities and their fixes from open-source software, in: Proceedings of
  the 17th International Conference on Predictive Models and Data Analytics in
  Software Engineering, 2021, pp. 30--39.

\bibitem{28}
J.~Fan, Y.~Li, S.~Wang, T.~N. Nguyen, Ac/c++ code vulnerability dataset with
  code changes and cve summaries, in: Proceedings of the 17th International
  Conference on Mining Software Repositories, 2020, pp. 508--512.

\end{thebibliography}

\end{document}